\documentclass[]{spie}  

\usepackage{amsmath,amsfonts,amssymb}
\usepackage{graphicx}
\usepackage[colorlinks=true, allcolors=blue]{hyperref}

\title{The Online Observation Quality System Software Architecture for the ASTRI Mini-Array Project}

\author[a]{N. Parmiggiani}
\author[a]{A. Bulgarelli} 
\author[a]{L. Baroncelli} 
\author[a]{A. Addis}
\author[a]{V. Fioretti} 
\author[a]{A. Di Piano} 
\author[b]{M. Capalbi}
\author[b]{O. Catalano} 
\author[a]{V. Conforti} 
\author[c,d]{M. Fiori}
\author[a]{F. Gianotti}
\author[e,f]{S. Iovenitti}
\author[g,h]{F. Lucarelli}
\author[b]{M. C. Maccarone} 
\author[b]{T. Mineo}
\author[g]{S. Lombardi} 
\author[a]{V. Pastore} 
\author[a]{F. Russo}
\author[b]{P. Sangiorgi} 
\author[i]{S. Scuderi} 
\author[l]{G. Tosti}
\author[a]{M. Trifoglio} 
\author[m]{L. Zampieri$^{\mathrm{d}}$, for the ASTRI Project}

\affil[a]{INAF/OAS Bologna, Via P. Gobetti 93/3, I-40129 Bologna, Italy.}

\affil[b]{INAF/IASF Palermo, Via Ugo La Malfa 153, I-90146 Palermo, Italy.}

\affil[c]{Universit\`{a} degli Studi di Padova, Via VIII Febbraio 2, I-35122 Padova, Italy}

\affil[d]{INAF/OA Padova, Vicolo Osservatorio 5, I-35122 Padova, Italy}

\affil[e]{INAF/OA Brera, via Brera 28, I-20121 Milano, Italy.}

\affil[f]{Universit\`{a} degli Studi di Milano, Dip.to di Fisica, Via Giovanni Celoria, 16, I-20133 Milano, Italy.}

\affil[g]{INAF/OAR Roma, Via di Frascati 33, I-00078 Monte Porzio Catone, Roma, Italy.}

\affil[h]{ASI/SSDC Roma, Via del Politecnico snc, I-00133 Roma, Italy.}

\affil[i]{INAF/IASF Milano, Via Alfonso Corti 12, I-20133 Milano, Italy.}

\affil[l]{Universit\`{a} degli Studi di Perugia, Dip.to di Fisica e Geologia, Via A. Pascoli, I-06123 Perugia, Italy}

\affil[m]{\protect\url{http://www.astri.inaf.it/en/library/}}

\authorinfo{Further author information: (Send correspondence to N. Parmiggiani)\\N. Parmiggiani: E-mail: nicolo.parmiggiani@inaf.it}

\begin{document} 
\maketitle

\begin{abstract}
The ASTRI Mini-Array is an international collaboration led by the Italian National Institute for Astrophysics. This project aims to construct and operate an array of nine Imaging Atmospheric Cherenkov Telescopes to study gamma-ray sources at very high energy (TeV) and perform stellar intensity interferometry observations. We describe the software architecture and the technologies used to implement the Online Observation Quality System (OOQS) for the ASTRI Mini-Array project. The OOQS aims to execute data quality checks on the data acquired in real-time by the Cherenkov cameras and intensity interferometry instruments, and provides feedback to both the Central Control System and the Operator about abnormal conditions detected. The OOQS can notify other sub-systems, triggering their reaction to promptly correct anomalies. The results from the data quality analyses (e.g. camera plots, histograms, tables, and more) are stored in the Quality Archive for further investigation and they are summarised in reports available to the Operator. Once the OOQS results are stored, the operator can visualize them using the Human Machine Interface. The OOQS is designed to manage the high data rate generated by the instruments (up to 4.5 GB/s) and received from the Array Data Acquisition System through the Kafka service. The data are serialized and deserialized during the transmission using the Avro framework. The Slurm workload scheduler executes the analyses exploiting key features such as parallel analyses and scalability.
\end{abstract}


\section{Introduction}\label{sec:intro}

The ASTRI Mini-Array\cite{2016JPhCS.718e2028P,Antonelli:2021ml} is a ground-based project led by INAF that aims to deploy and operate nine innovative dual-mirror Imaging Atmospheric Cherenkov Telescopes (IACTs) \cite{Krennrich_2009} to study gamma-ray sources emitting at very high energy in the TeV spectral band and to perform intensity interferometry observations. Each telescope is equipped with a new ASTRICAM Silicon photomultiplier Cherenkov camera with its fast read-out electronics specifically designed for this purpose,\cite{2018SPIE10702E..37C} and managed by the Cherenkov Camera Software Supervisor \cite{Corpora:2021Oh,Corpora2022}. In addition to Cherenkov observations, the ASTRI Mini-Array performs intensity interferometry observations of a selected list of bright sources. For this purpose, each telescope is equipped with a Stellar Intensity Interferometry Instrument \cite{Zampieri2022} (SI3 or SI$^3$). The nine telescopes are under installation at the Teide Astronomical Observatory, operated by the Instituto de Astrofisica De Canarias (IAC), on Mount Teide (2400 m a.s.l.) in Tenerife (Canary Islands, Spain). INAF has a host agreement with the IAC to operate them. 

The ASTRI Mini-Array can be operated remotely, and no human presence is foreseen on-site during observations. A data center is being installed on-site to host several systems that operate the telescopes and manage the data acquired by the instruments. The Array Data Acquisition System\cite{Conforti2022} (ADAS), deployed in the on-site data center, acquires the raw data from the telescopes to locally store them and, in parallel, send the data to the off-site data center in Rome. In addition, the ADAS implements an interface\cite{Pastore2022} with the Online Observation Quality System (OOQS) to send the data acquired. The OOQS performs a quick-look analysis during the observations.

This contribution describes the software architecture and design of the OOQS. The OOQS aims to perform the data quality check in real-time during the observations and inform the Central Control System or the Alarm System if abnormal conditions are detected within the data. The fast analyses of the data quality allow the ASTRI Mini-Array sub-systems to take automated corrective actions and notify the Operator if the data acquired by the telescopes do not satisfy the quality requirements.

The rest of the manuscript is organized as follows. Section \ref{sec:use_cases} introduces the use cases and software requirements of the OOQS. Section \ref{sec:architecture} describes the software architecture of the OOQS and its software components. Section \ref{sec:dev_plan} reports the process and technologies used during the software development and describes the deployment schema of the OOQS in the on-site data center. Section \ref{sec:conclusion} presents the conclusions.

\section{Use Cases and Requirements} \label{sec:use_cases}

\begin{figure*}[!htb]
	\centering
	  \includegraphics[width=\textwidth]{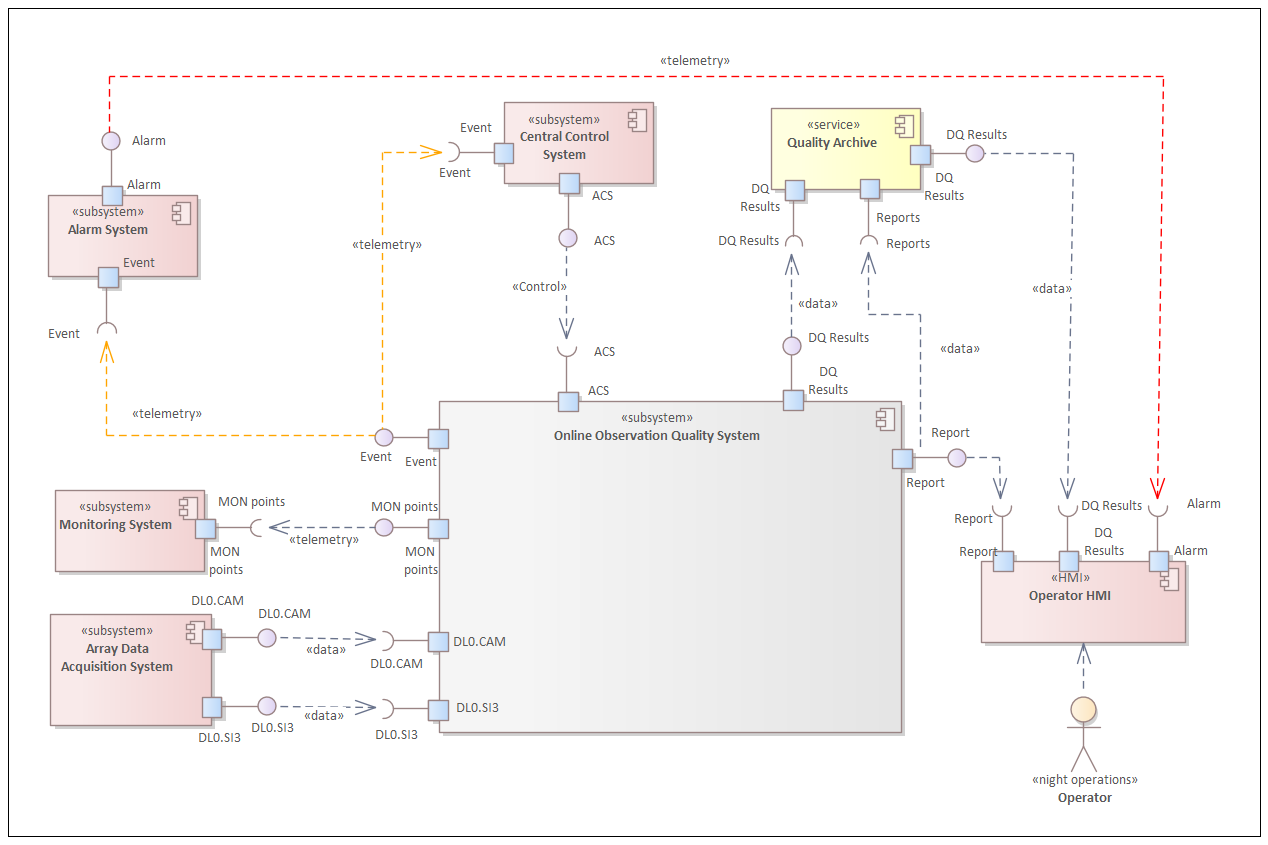}
	\caption{The system context diagram of the OOQS. The Central Control System supervises the OOQS components. The OOQS receives data acquired by the telescopes from the ADAS through Kafka, performs the quality checks, and stores the results in the Quality Archive. The Operator can visualize the results through the Operator HMI. The OOQS sends the abnormal conditions to the Central Control System and Alarm System (orange arrows). The Monitoring System receives monitoring information about the software status.}
	\label{fig:sys_context}
\end{figure*}

Figure \ref{fig:sys_context} shows the system context in which the OOQS shall execute the data quality checks and the interfaces with other sub-systems. The Central Control system control the OOQS lifecycle, starting and stopping it. The ADAS sends to the OOQS the raw data (DL0) acquired during Cherenkov or intensity interferometry observations. The OOQS performs a list of data quality checks on the data received and stores the results of these analyses in the Quality Archive. The Operator can visualize the data quality results using the Operator Human Machine Interface (HMI). The OOQS sends telemetry data to the Alarm System and Central Control System to notify possible abnormal conditions detected during the quality check analysis. When the Alarm system receives the notification about an abnormal condition, it evaluates the possibility of sending an alarm to the Operator through the HMI. The Monitoring System \cite{Costa:2021c8,Incardona2022} receives from the OOQS information about the status of the software components.

During the first phase of the design process, we defined the use cases and the requirements that the OOQS have to satisfy, starting from the top-level use case and requirements of the whole ASTRI Mini-Array on-site software system \cite{Bulgarelli2022}. During this process, we considered the experience and the know-how that the team has acquired in other projects: the ASTRI Horn \cite{10.1117/12.2530896} prototype, and the AGILE \cite{2008NIMPA.588...52T} space mission. During the definition of the use cases, we collaborated with domain experts to collect information on the quality checks that the OOQS shall perform to evaluate the data acquired by the telescopes. 
    
The use cases must follow the two operative modes of the OOQS, one during the Cherenkov observations and another during SI$^3$ data acquisition. For this reason, we split the OOQS functionalities between the two different observing modes. This decision led to different software components described in the following sections, specialized for one or other observing mode. 
    
Starting from the functionalities extracted from the use cases, we defined the list of the main functional components that constitute the OOQS. These components are described in detail in Section \ref{sec:architecture} and are: 
    
\begin{itemize}
    \item OOQS Master: the component that manages the lifecycle of OOQS components.
    \item OOQS Manager: the component that manages the software components that execute the quality checks during Cherenkov and intensity interferometry observations and is interfaced with the Central Control System, the Alarm System, and the Monitoring System.
    \item Cherenkov Camera Data Quality Checker (CCDQC): the component that performs the data quality check analysis during the Cherenkov observations.
    \item SI$^3$ Data Quality Checker (SI3DQC): the component that performs data quality check analysis during the intensity interferometry observations.
    \item Quality Archive: the archive deployed in the on-site data center that stores all the results of the OOQS. 
\end{itemize}

The use cases and the software components are described in detail in Section \ref{subsec:master_manager}, Section \ref{subsec:ccdqc}, and Section \ref{subsec:SI3DQC}.
    
The next step was to define the software requirements for each software component. The requirements can be of two categories: functional or performance requirements. The functional requirements define \textit{what} action each component shall complete, while the performance requirements define \textit{how} those actions must be completed. During the OOQS verification and validation process, the quality assurance team check that all the requirements are satisfied.

\section{OOQS Software Architecture and Design}\label{sec:architecture}

\begin{figure*}[!htb]
	\centering
	  \includegraphics[width=\textwidth]{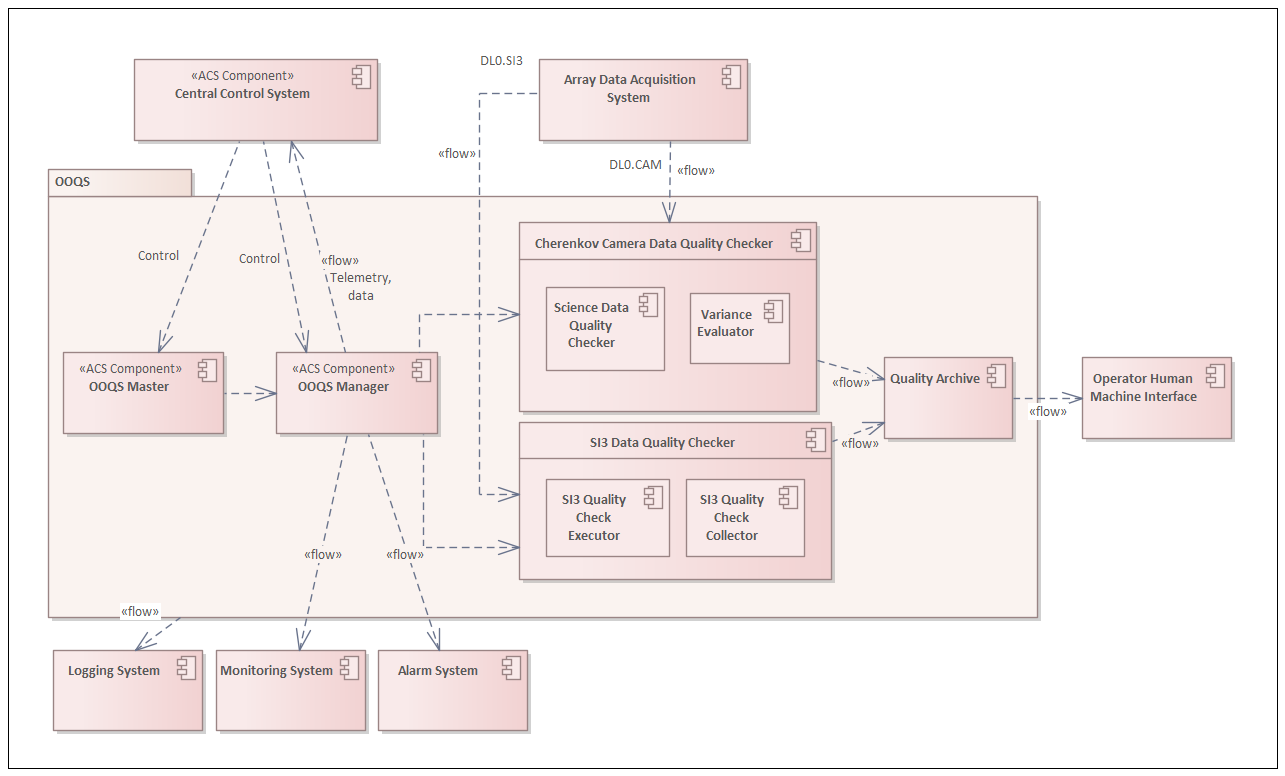}
	\caption{Internal composition of the OOQS. It can instantiate the Cherenkov Camera Data Quality Checker or the SI$^3$ Data Quality Checker during the Cherenkov or intensity interferometry observations respectively. These components have sub-components specialized to analyze the different data types acquired during the observations.}
	\label{fig:scada_arch}
\end{figure*}

Figure \ref{fig:scada_arch} shows the internal composition of the OOQS with the functional software components and their sub-components. The OOQS is part of the ASTRI Mini-Array Supervisory Control and Data Acquisition system (SCADA). SCADA is the system that shall manage the startup, shutdown, configuration, and control of all site assemblies and sub-systems to collect monitoring points, manage alarms raised by any assembly, check the health status of all systems, and acquire scientific data. All SCADA sub-systems, from which the OOQS, are deployed in the on-site data center to have a fast and reliable connection.

We implement the OOQS Master and the OOQS Manager with the Alma Common Software\cite{2002SPIE.4848...43C} (ACS). ACS is a framework used to develop distributed software and supports several programming languages: Python, C++, and Java. It collects common patterns and implements them in ACS components. The ASTRI Mini-Array team uses ACS for the development of SCADA sub-systems to standardize the software architecture and the interfaces between the different software systems. 

The CCDQC and the SI3DQC are software components not developed using ACS and run outside the ACS environment. The CCDQC comprises two software components to analyze different data acquired by the nine Cherenkov cameras, described in detail in Section \ref{subsec:ccdqc}. The SI3DQC has a different internal structure due to the high data throughput generated by the SI$^3$, described in Section \ref{subsec:SI3DQC}. It is composed of the SI$^3$ Quality Check Executor and a the SI$^3$ Quality Check Collector.

\begin{figure*}[!htb]
	\centering
	  \includegraphics[width=0.7\textwidth]{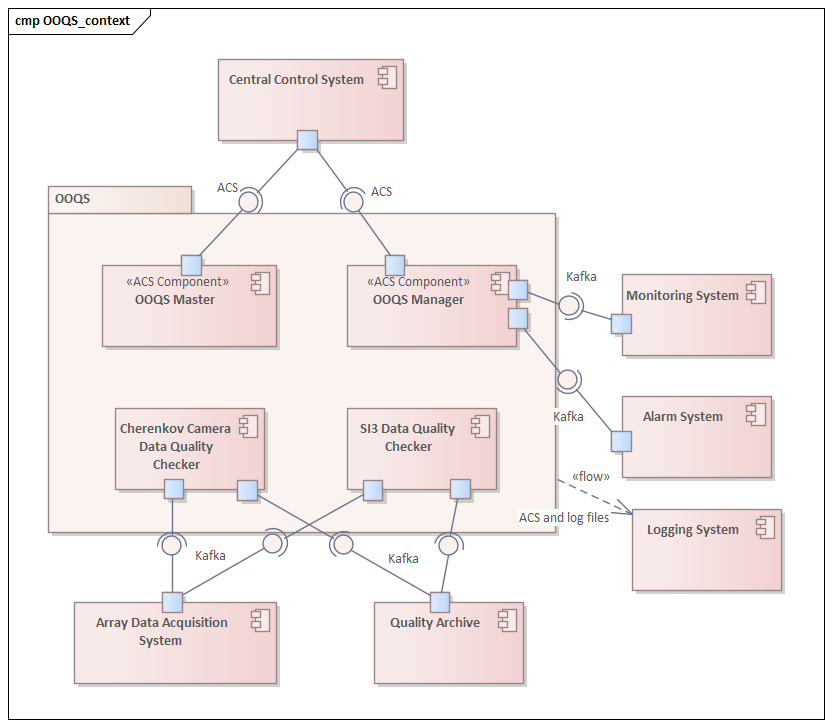}
	\caption{Interfaces between the OOQS and other sub-systems. The Interfaces with the Central Control System are managed using the ACS framework. The ADAS sends input data to the OOQS using Kafka. The OOQS sends data to the Monitoring System, the Alarm System and the Quality Archive through Kafka.}
	\label{fig:interfaces}
\end{figure*}

Figure \ref{fig:interfaces} shows the interface protocols used between the OOQS and other SCADA sub-systems. These interfaces are described below and exploit the features of frameworks and services designed for the Internet Of Things and Big Data context.

The CCDQC and SI3DQC components execute the quality checks on the data received from the ADAS system through the Apache Kafka\footnote{https://kafka.apache.org} service. Kafka is an open-source distributed event streaming platform. Kafka is used to implement high-performance pipelines, streaming analytics, and mission-critical applications. The SCADA developers decided to exploit the Kafka features such as the high throughput, the scalability, and the high availability. Kafka can store key-value messages generated by an arbitrary number of producers. The messages are part of different topics, and each topic is composed of many partitions. The consumer process can read the messages from the desired topics. Kafka will be deployed in a cluster of multiple servers (brokers); Kafka partitions are therefore distributed and replicated across the cluster in order to improve the fault tolerance and scalability. In addition, Kafka manages the retention period of each message based on the time, or the space bound specified in the configuration. Kafka can delete these records to free storage space when the records are older than the configured retention time or if the space used exceeds the limit. 

The ADAS implements Kafka producers that send the data acquired by the telescopes to the Kafka consumer implemented by the OOQS. The two systems operate in different physical servers (Section \ref{sec:dev_plan}), and they use Kafka to share data.

\begin{figure*}[!htb]
	\centering
	  \includegraphics[width=0.8\textwidth]{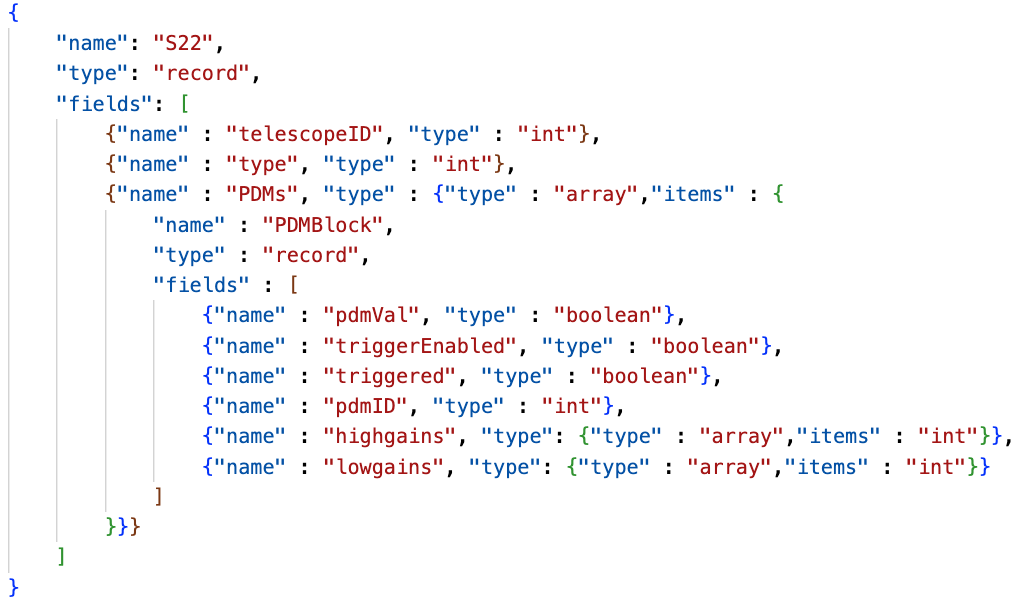}
	\caption{Part of the Avro schema used to serialize the S(2,2) packet type sent by the ADAS to the OOQS through Kafka.}
	\label{fig:avro_schema}
\end{figure*}

The Kafka producer serializes the data using Apache Avro\footnote{https://avro.apache.org} to fix the messages size and reduce the overall data rate. Avro is a serialization framework that serializes data in a binary format and uses JSON schema (Figure \ref{fig:avro_schema}) to structure the data for the encoding. For this reason, the schema is an interface between the ADAS and OOQS. The ADASS encodes the data packet by packet before sending them through Kafka, and the OOQS decodes them to start the quality check analyses.

We decided to manage the software components that execute the data quality checks and run outside the ACS framework with the Slurm\footnote{http://slurm.schedmd.com} workload manager. Slurm can schedule and execute multiple processes in parallel to optimize the resources. In addition, Slurm can run in a cluster composed of several workers and one controller to balance the job execution and manage the failures of the workers. Slurm can also manage the priority between processes. When a high-priority job is scheduled and the resources are not enough,  Slurm suspends the low-priority job to execute the high-priority one. When the high-priority job ends, Slurm resumes the low-priority job. Slurm can automatically replace failed jobs with a new ones.

\subsection{Master and Manager Components}\label{subsec:master_manager}

The OOQS Master is implemented as an ACS Master Component. It aims to control the lifecycle of other OOQS ACS components. The OOQS Master has an interface with the Central Control System. The Central Control System sends requests to the OOQS Master for the startup and the shutdown of all the OOQS components. In addition, the OOQS Master monitors the status of the OOQS components during the operations and manages the exceptions raised by the software systems. The ASTRI Mini-Array has nine OOQS Master components, one for each telescope. 

The OOQS Manager is an ACS component. It supervises the software that executes the quality checks on the data received by the ADAS. There are nine OOQS Manager components, one for each of the nine telescopes. Every OOQS Master supervises one OOQS Master. The OOQS Master starts, stops, and controls the supervised OOQS Manager. The OOQS Manager sends the abnormal conditions found during the analyses to the Central Control System and the Alarm system. It also has an interface with the Monitoring System to send the status of the software components that execute the quality checks (e.g., the number of processes, event rate, processing time, and more). The OOQS Manager can instantiate two types of data quality software: (i) the CCDQC to perform quality checks on the data acquired during the Cherenkov observations and (ii) the SI3DQC to perform different analyses during the intensity interferometry data acquisition. The ASTRI Mini-Array cannot observe simultaneously with the Cherenkov camera and the SI$^3$. For this reason, the OOQS Manager instantiates the appropriate software based on the observation type.  

\subsection{Cherenkov Camera Data Quality Checker}\label{subsec:ccdqc}

\begin{figure*}[!htb]
	\centering
	  \includegraphics[width=\textwidth]{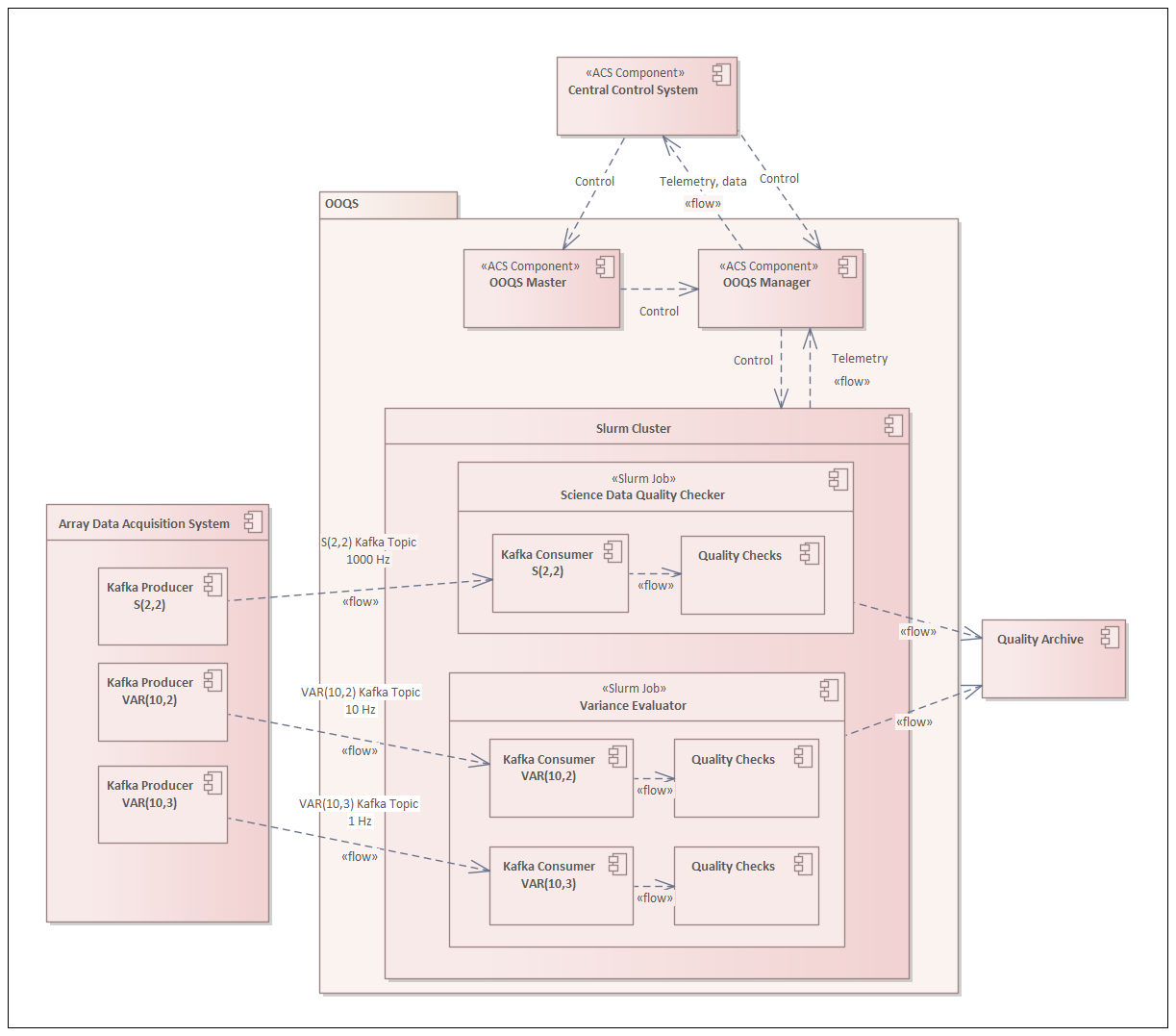}
	\caption{High-level schema of the Cherenkov Camera Data Quality Checker component. The Central Control System instantiates this component when a Cherenkov observation is started. The ADAS sends the data acquired by the telescope to the OOQS through three Kakfa topics. The OOQS instantiate Slurm jobs to execute the Science Data Quality Checker on the S(2,2) packet and the Variance Evaluator on VAR packets. The results are stored in the Quality archive and abnormal conditions are sent back to the Central Control System as telemetry information.}
	\label{fig:cherenkov_workflow}
\end{figure*}

Figure \ref{fig:cherenkov_workflow} shows the workflow of the CCDQC. This software component aims to perform data quality checks on the data received from the ADAS system during Cherenkov observations. The Central Control Systems send a request to the OOQS Master to activate the CCDQC when a new Cherenkov observation is executed. The OOQS Master asks the OOQS Manager to activate this software. Once the CCDQC is started, it submits to the Slurm Cluster the processes to manage the workload and optimize the computational resources. The ADAS sends the DL0 to the OOQS through the Kafka service described in Sect. \ref{sec:architecture}. 

The CCDQC has two software components: the Science Data Quality Checker and the Variance Evaluator. These two components analyze the data packets generated by the Cherenkov camera: the scientific data packets, called S(2,2), and two variance data packets, the VAR(10,2) and VAR(10,3). The ADAS sends these data through different Kafka topics, and the OOQS implements the same number of consumers to read and analyze the data packets (Figure \ref{fig:cherenkov_workflow}). 

The S(2,2) packet has a size of about 13 kB. It contains the high gain and low gain Analog-to-Digital Converter (ADC) values for the 64 pixels of the 37 Photon Detection Modules (PDMs) constituting each Cherenkov camera, the time of the pixel triggers, temperatures, and other information about the acquisition context. The maximum event rate for each camera is 1000 Hz. The maximum data rate for each telescope is about 13 MB/s, 117 MB/s considering the nine telescopes. The VAR packet has a size of about 9.5 kB, but the event rate is between 1 and 10 Hz; thus, they don't constitute a challenging data rate. The cameras produce the VAR(10,2) and VAR(10,3) packets by randomly sampling the pixels' values without a trigger. The variance of these values is calculated separately for the high-gain (VAR(10,2)) and low-gain (VAR(10,3)) chains. The two VAR packets have the same data structure.

The main data quality checks performed on the S(2,2) packets are:
\begin{enumerate}
	\item calculate the histograms of the trigger number for each camera and each camera's PDM;
	\item calculate the histogram of the times between two consecutive triggers for each camera;
 	\item check that the histogram values are inside a predefined range;
	\item check that the pixel ADC values of each Cherenkov camera are inside a predefined range;
	\item sample the data to obtain one camera image per second.
	\item execute checks 1, 2 and 4 with the calibrated data obtained with predefined calibration coefficients;
\end{enumerate} 

We implement the data quality checks listed above using the rta-dq-lib\cite{2021arXiv210508648B}, a framework developed to execute quality checks and optimized with high performances computing technologies. This framework can be configured using XML files with several types of analyses. The execution of the analyses is demanded to the rta-pipe-dq, another framework that instantiates several pipelines in parallel to distribute the workload in the Slurm Cluster. 

The main data quality checks performed on the VAR(10,2) and VAR(10,3) are:
\begin{enumerate}
	\item aggregate all the VAR data from the start of the observation for each camera;
	\item calculate the ratio between the high-gain and low-gain of each camera PDM; 
	\item check if the pointing deviation and the point spread function (PSF) size are inside the nominal range;
	\item sample the data to obtain one camera image per second.
\end{enumerate} 

In this context, the main purpose of the VAR analysis is to check if a pointing correction is necessary during observations. To this end, a specific technique has been developed on purpose \cite{IOVENITTI_CAMERA_ALIGNMENT}, and the eventual corrections can be applied in real-time or during the following observing run.

\subsection{SI$^3$ Data Quality Checker}\label{subsec:SI3DQC}

\begin{figure*}[!htb]
	\centering
	  \includegraphics[width=\textwidth]{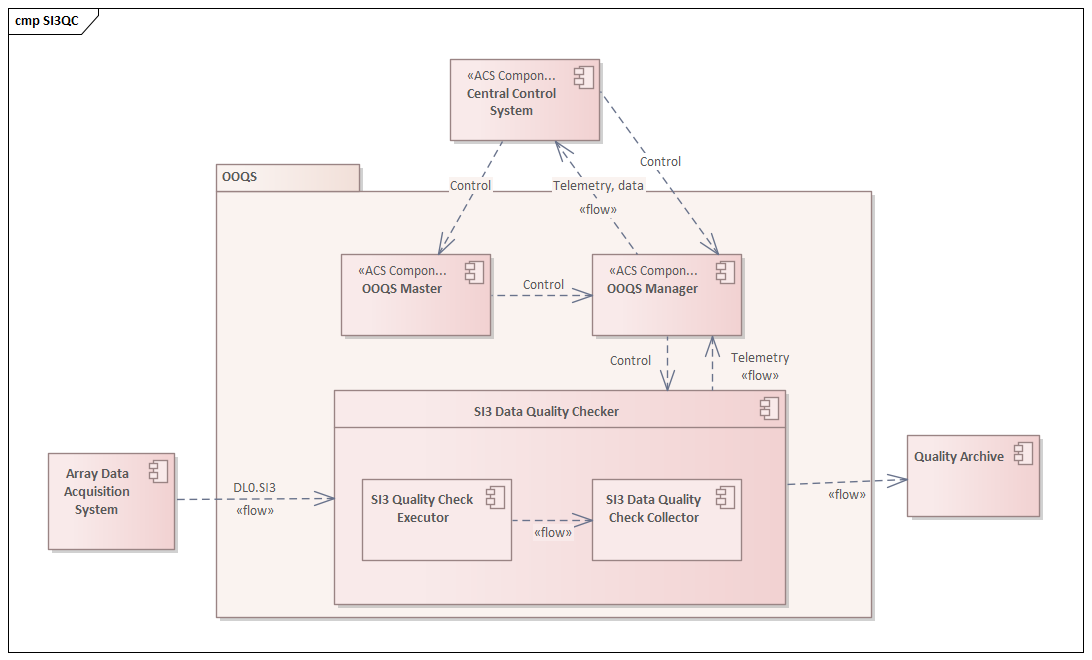}
	\caption{High-level schema of the SI$^3$ Data Quality Checker component. The Central Control System instantiates this component when an intensity interferometry observation is started. The ADAS sends the data acquired by the telescope to the OOQS to perform quality checks and detect abnormal conditions. The OOQS sends the abnormal conditions to the Central Control System as telemetry data and stores the results in the Quality Archive.}
	\label{fig:si3_workflow}
\end{figure*}

This software component aims to perform the online data quality check on the data acquired by the SI$^3$. The ASTRI Mini-Array has one SI3DQC for each telescope. The intensity interferometry observations produce a data rate higher than those obtained during the Cherenkov acquisition. Each SI$^3$ generates $100\times10^6$ events/second. The events have a size of 5 bytes, and the data rate for each telescope is 500 MB/s. The event rate for the full array of nine telescopes is $900\times10^6$ events/second, and the data rate is about 4,5 GB/s. We designed this component (Figure \ref{fig:si3_workflow}) in a different way to manage this demanding throughput. The SI3DQC is a distributed software between different servers. The OOQS  Manager controls the data quality analyses with two components, the SI$^3$ Quality Checker Executor and the SI$^3$ Quality Check Collector. The first one performs the quality checks. We are evaluating different configurations to avoid data transfer (e.g., running the Executor directly in the Camera Server where ADAS acquires data). The second component receives back the results of the quality checks, performs some additional inspections, and detects abnormal conditions that are sent as telemetry data to the Central Control System. The results of the analyses are stored in the Quality Archive.

The main data quality checks performed on the data acquired during the SI$^3$ observations are: 
\begin{enumerate}
	\item extract the event rate for each telescope from the raw data stream;
	\item check the pointing accuracy using the event rate acquired by the four quadrants of each SI$^3$ detector;
	\item select a random sub-sample of events from the data stream to check the content of the data packet. The OOQS cannot execute the checks on all packets due to the data rate (4.5 GB/s). It is enough to check a sample of packets to detect this kind of anomaly.
\end{enumerate}

\section{Software Implementation and Deployment} \label{sec:dev_plan}

We are developing the OOQS software following the ASTRI Mini-Array Software Development Plan guidelines that indicate common patterns and rules. We manage the software version control using the Git\footnote{https://git-scm.com} software and store the repository in the GitLab\footnote{https://about.gitlab.com} instance hosted by the INAF ICT department. We implemented a Continuous Integration (CI) pipeline following the DevOps principles \cite{2019arXiv190905409L} and using the GitLab built-in CI features to build and test the OOQS software when a new version of the code is uploaded. The CI pipeline executes a Docker\footnote{https://www.docker.com/} container where all the required software and libraries are installed. The execution of the pipelines is performed in a remote server configured as a GitLab Runner\footnote{https://docs.gitlab.com/runner/}. In addition, we implemented a pipeline for the static code analysis using SonarQube\footnote{https://www.sonarqube.org/}. 

The OOQS software and dependencies (e.g., external packages and services) are managed using Docker to exploit the key features of this service. Docker allows the developer to package and run an application in an isolated environment called the container. Thanks to the isolation and the security granted by this system, it is possible to run several containers simultaneously on the same machine, avoiding conflicts. The containers include all the software dependencies required, and there is no need to use the software installed on the host machine. The containers can be shared and deployed in different environments, increasing the system's flexibility. The containerization technique has several advantages: 

\begin{itemize}
    \item all the software dependencies can be installed in a single place, the container;
    \item the configuration to obtain the environment can be managed with a version control system;
    \item the containers are isolated, and it is possible to run several containers in parallel on the same machine;
    \item it is possible to obtain reproducible environment and results.
    \item the containers can be used to implement a CI workflow.
\end{itemize}

\begin{figure*}[!htb]
	\centering
	  \includegraphics[width=\textwidth]{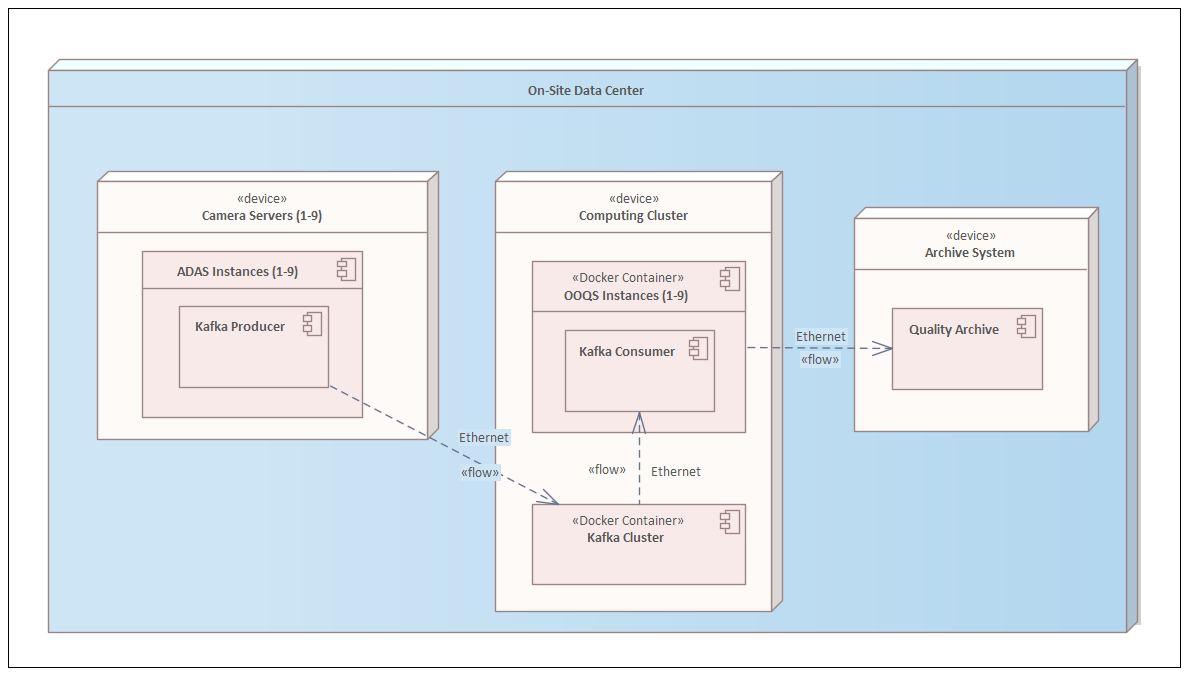}
	\caption{OOQS deployment in the on-site data center. The nine OOQS instances are deployed in the on-site Computing Cluster using Docker containers. The nine ADAS instances, hosted in nine Camera Servers, send the data to the OOQS intances through Kafka. Finally, the OOQS instances send the results of their analyses to the Quality Archive part of the Archive System.}
	\label{fig:phys_context}
\end{figure*}

Figure \ref{fig:phys_context} shows the deployment schema of the OOQS inside the ASTRI Mini-Array on-site data center. The architecture foresees nine Camera Servers hosting nine ADAS instances, installed on the bare-metal, to acquire data from the nine telescopes through a point-to-point direct connection between the instruments and the Camera Servers and send it to nine OOQS instances through Kafka. The OOQS instances, with all their software components, are deployed as Docker containers in the Computing Cluster composed of multiple servers shared with other sub-systems (Monitoring System, Alarm System, Archive, etc.). The results obtained by the OOQS are sent to the Quality Archive part of the on-site Archive System.
 
\section{Conclusions}\label{sec:conclusion}

We described in this manuscript the software architecture and the design of the OOQS system for the ASTRI Mini-Array project. This software aims to perform real-time data quality checks on the data acquired by the ADAS from the nine telescopes and sent to the OOQS through Kafka.

The OOQS analyses the data acquired during the Cherenkov and the stellar intensity interferometry observations. The data packets produced during these two observing modes are different. For this reason, the OOQS instantiates the software designed to perform the quality checks on the data acquired by the Cherenkov cameras or the SI$^3$. Slurm is the workload manager that executes the analyses, exploiting key features such as failure management and load balancing. The data quality checks and the performance requirements for the two observing modes are different. In the case of the intensity interferometry observations, the data rate can reach a maximum of 4.5 GB/s for the whole array, and this requirement needs a dedicated solution. On the other hand, the OOQS must be flexible enough to execute many analyses on different data packets acquired during Cherenkov observations. 

The software and dependencies are installed into a Docker container to grant the environment reproducibility and deploy the OOQS in the on-site data center in a cluster of servers shared with other sub-systems. The containerization is essential for the isolation of the OOQS in this deployment context. In addition, the CI pipeline, implemented during the development phase, uses the Docker containers to execute and test the software.

The OOQS is a critical system because it can detect abnormal conditions and notify the Central Control System and the Alarm System for a fast reaction. In addition, the Operator can visualize the OOQS results stored in the Quality Archive through the HMI during the supervision of the observations and take corrective actions if needed.

\acknowledgments 
 
This work was conducted in the context of the ASTRI Project thanks to the support of the Italian Ministry of University and Research (MUR) as well as the Ministry for Economic Development (MISE) with funds specifically assigned to the Italian National Institute of Astrophysics (INAF). We acknowledge support from the Brazilian Funding Agency FAPESP (Grant 2013/10559-5) and from the South African Department of Science and Technology through Funding Agreement 0227/2014 for the South African Gamma-Ray Astronomy Programme. IAC is supported by the Spanish Ministry of Science and Innovation (MICIU). This work has also been partially supported by H2020-ASTERICS, a project funded by the European Commission Framework Programme Horizon 2020 Research and Innovation action under grant agreement n. 653477. The ASTRI project is becoming a reality thanks to Giovanni “Nanni” Bignami, Nicol\'{o} “Nichi” D’Amico two outstanding scientists who, in their capability of INAF Presidents, provided continuous support and invaluable guidance. While Nanni was instrumental to start the ASTRI telescope, Nichi transformed it into the Mini Array in Tenerife. Now the project is being built owing to the unfaltering support of Marco Tavani, the current INAF President. Paolo Vettolani and Filippo Zerbi, the past and current INAF Science Directors, as well as Massimo Cappi, the Coordinator of the High Energy branch of INAF, have been also very supportive to our work. We are very grateful to all of them. Nanni and Nichi, unfortunately, passed away but their vision is still guiding us. This article has gone through the internal ASTRI review process.

\bibliography{report} 

\begin{thebibliography}{10}

\bibitem{2016JPhCS.718e2028P}
{Pareschi}, G., {Bonnoli}, G., {Vercellone}, S., {ASTRI Collaboration}, and
  {CTA Consortium}, ``{The mini-array of ASTRI SST-2M telescopes, precursors
  for the Cherenkov Telescope Array},'' in [{\em Journal of Physics Conference
  Series}{\nolinebreak\hspace{0.1em}]},  {\em Journal of Physics Conference
  Series} {\bf 718},  052028 (May 2016).

\bibitem{Antonelli:2021ml}
Antonelli, L.~A., ``{The ASTRI Mini-Array at Teide Observatory},'' in [{\em
  Proceedings of 37th International Cosmic Ray Conference {\textemdash}
  PoS(ICRC2021)}{\nolinebreak\hspace{0.1em}]},   {\bf 395},  897 (2021).

\bibitem{Krennrich_2009}
Krennrich, F., ``Gamma ray astronomy with atmospheric cherenkov telescopes: the
  future,'' {\em New Journal of Physics}~{\bf 11},  115008 (nov 2009).

\bibitem{2018SPIE10702E..37C}
{Catalano}, O., {Capalbi}, M., {Gargano}, C., {Giarrusso}, S., {Impiombato},
  D., et~al., ``{The ASTRI camera for the Cherenkov Telescope Array},'' in
  [{\em Ground-based and Airborne Instrumentation for Astronomy
  VII}{\nolinebreak\hspace{0.1em}]},  {Evans}, C.~J., {Simard}, L., and
  {Takami}, H., eds., {\em Society of Photo-Optical Instrumentation Engineers
  (SPIE) Conference Series} {\bf 10702},  1070237 (July 2018).

\bibitem{Corpora:2021Oh}
Corpora, M., Grillo, A., Sangiorgi, P., Capalbi, M., Catalano, O., et~al.,
  ``{The Architecture of ASTRI Mini Array Cherenkov Camera Software
  Supervisor},'' in [{\em Proceedings of 37th International Cosmic Ray
  Conference {\textemdash} PoS(ICRC2021)}{\nolinebreak\hspace{0.1em}]},   {\bf
  395},  691 (2021).

\bibitem{Corpora2022}
Corpora, M., Grillo, A., Sangiorgi, P., Capalbi, M., Catalano, O., et~al.,
  ``{Design and development of the Supervisor Software Component for the ASTRI
  Mini-Array Cherenkov Camera},'' in [{\em Software and Cyberinfrastructure for
  Astronomy VII}{\nolinebreak\hspace{0.1em}]},  Jorge~Ibsen, G.~C., ed.,  {\bf
  Proc. of SPIE 12189} (2022).

\bibitem{Zampieri2022}
Zampieri, L., Bonanno, G., Bruno, P., Gargano, C., Lessio, L., et~al., ``{A
  Stellar Intensity Interferometry Instrument for the ASTRI Mini-Array
  telescopes},'' in [{\em Optical and Infrared Interferometry and Imaging
  VIII}{\nolinebreak\hspace{0.1em}]},  Antoine~Mérand, Stephanie~Sallum, J.
  S.-B., ed.,  {\bf Proc. of SPIE 12183} (2022).

\bibitem{Conforti2022}
Conforti, V., Gianotti, F., Pastore, V., Trifoglio, M., Bulgarelli, A., et~al.,
  ``{The Array Data Acquisition System software architecture of the ASTRI
  Mini-Array Project},'' in [{\em Software and Cyberinfrastructure for
  Astronomy VII}{\nolinebreak\hspace{0.1em}]},  Jorge~Ibsen, G.~C., ed.,  {\bf
  Proc. of SPIE 12189} (2022).

\bibitem{Pastore2022}
Pastore, V., Conforti, V., Gianotti, F., Bulgarelli, A., and Parmiggiani, N.,
  ``{Array Data Acquisition System interface for online distribution of
  acquired data in the ASTRI Mini-Array project},'' in [{\em Software and
  Cyberinfrastructure for Astronomy VII}{\nolinebreak\hspace{0.1em}]},
  Jorge~Ibsen, G.~C., ed.,  {\bf Proc. of SPIE 12189} (2022).

\bibitem{Costa:2021c8}
Costa, A., Munari, K., Incardona, F., Bruno, P.~G., Germani, S., et~al., ``{The
  Monitoring, Logging, and Alarm system for the Cherenkov Telescope Array},''
  in [{\em Proceedings of 37th International Cosmic Ray Conference
  {\textemdash} PoS(ICRC2021)}{\nolinebreak\hspace{0.1em}]},   {\bf 395},  700
  (2021).

\bibitem{Incardona2022}
Incardona, F., Costa, A., Munari, K., Gambadoro, S., Bruno, P., et~al., ``{The
  Monitoring Logging and Alarm System of the ASTRI Mini-Array gamma-ray
  air-Cherenkov experiment at the Observatorio del Teide},'' in [{\em Software
  and Cyberinfrastructure for Astronomy VII}{\nolinebreak\hspace{0.1em}]},
  Jorge~Ibsen, G.~C., ed.,  {\bf Proc. of SPIE 12189} (2022).

\bibitem{Bulgarelli2022}
Bulgarelli, A., Lucarelli, F., Tosti, G., Conforti, V., Parmiggiani, N.,
  et~al., ``{The Software Architecture and development approach for the ASTRI
  Mini-Array gamma-ray air-Cherenkov experiment at the Observatorio del
  Teide},'' in [{\em Software and Cyberinfrastructure for Astronomy
  VII}{\nolinebreak\hspace{0.1em}]},  Jorge~Ibsen, G.~C., ed.,  {\bf Proc. of
  SPIE 12189} (2022).

\bibitem{10.1117/12.2530896}
Giro, E., Canestrari, R., Bruno, P., Catalano, O., Fugazza, D., et~al., ``{The
  ASTRI-Horn telescope validation toward the production of the ASTRI
  Mini-Array: a proposed pathfinder for the Cherenkov Telescope Array},'' in
  [{\em Optics for EUV, X-Ray, and Gamma-Ray Astronomy
  IX}{\nolinebreak\hspace{0.1em}]},  O'Dell, S.~L. and Pareschi, G., eds.,
  {\bf 11119},  446 -- 460, International Society for Optics and Photonics,
  SPIE (2019).

\bibitem{2008NIMPA.588...52T}
{Tavani}, M. et~al., ``{The AGILE space mission},'' {\em Nuclear Instruments
  and Methods in Physics Research A}~{\bf 588},  52--62 (Apr. 2008).

\bibitem{2002SPIE.4848...43C}
{Chiozzi}, G., {Gustafsson}, B., {Jeram}, B., {Plesko}, M., {Sekoranja}, M.,
  et~al., ``{CORBA-based Common Software for the ALMA project},'' in [{\em
  Advanced Telescope and Instrumentation Control Software
  II}{\nolinebreak\hspace{0.1em}]},  {Lewis}, H., ed., {\em Society of
  Photo-Optical Instrumentation Engineers (SPIE) Conference Series} {\bf 4848},
   43--54 (Dec. 2002).

\bibitem{2021arXiv210508648B}
{Baroncelli}, L., {Bulgarelli}, A., {Parmiggiani}, N., {Fioretti}, V., {Addis},
  A., et~al., ``{rta-dq-lib: a software library to perform online data quality
  analysis of scientific data},'' {\em arXiv e-prints} ,  arXiv:2105.08648 (May
  2021).

\bibitem{IOVENITTI_CAMERA_ALIGNMENT}
Iovenitti, S., Sironi, G., Giro, E., Segreto, A., Catalano, O., et~al.,
  ``Assessment of the {Cherenkov} camera alignment through {Variance} images
  for the {ASTRI} telescope,'' {\em Experimental Astronomy}~{\bf 53},  117--132
  (Nov. 2021).

\bibitem{2019arXiv190905409L}
{Leite}, L., {Rocha}, C., {Kon}, F., {Milojicic}, D., and {Meirelles}, P., ``{A
  Survey of DevOps Concepts and Challenges},'' {\em arXiv e-prints} ,
  arXiv:1909.05409 (Sept. 2019).

\end{thebibliography}
\bibliographystyle{spiebib} 

\end{document}